\documentclass{article}

\usepackage{arxiv}

\usepackage[utf8]{inputenc} 
\usepackage[T1]{fontenc}    
\usepackage{hyperref}       
\usepackage{url}            
\usepackage{booktabs}       
\usepackage{amsfonts}       
\usepackage{nicefrac}       
\usepackage{microtype}      
\usepackage{lipsum}
\usepackage{graphicx}
\usepackage{xcolor}
\usepackage{tipa}

\title{The voice of COVID-19: \\ Acoustic correlates of infection}

\author{
  Katrin D.~Bartl-Pokorny\thanks{Equal contribution. Correspondence to: \texttt{firstname.lastname@uni-a.de}}~~\&~Florian B. Pokorny\footnotemark[1] \\
  EIHW -- Chair of Embedded Intelligence for Health Care and Wellbeing, University of Augsburg, Augsburg, Germany   \& \\
 Division of Phoniatrics, Medical University of Graz, Graz, Austria \\
  \And
  Anton Batliner, Shahin Amiriparian, Anastasia Semertzidou \\
  EIHW -- Chair of Embedded Intelligence for Health Care and Wellbeing, University of Augsburg, Augsburg, Germany \\
  \And
  Florian Eyben \\ 
  audEERING GmbH, Gilching, Germany \\
  \And
  Elena Kramer, Florian Schmidt, Rainer Sch\"onweiler \\
  Department of Otorhinolaryngology, Phoniatrics and Paediatric Audiology \\
  University Hospital of Schleswig-Holstein, Campus Lübeck, Lübeck, Germany \\
  \And
  Markus Wehler\\
  Department of Emergency Medicine and Medicine IV, University Medical Center Augsburg, Augsburg, Germany \\
  \And
  Bj\"orn W.~Schuller \\
  EIHW -- Chair of Embedded Intelligence for Health Care and Wellbeing, University of Augsburg, Augsburg, Germany  \& \\
  GLAM -- Group on Language, Audio, \& Music, Imperial College London, London, United Kingdom
}

\begin{document}
\maketitle

\begin{abstract}
COVID-19 is a global health crisis that has been affecting many aspects of our daily lives throughout the past year. The symptomatology of COVID-19 is heterogeneous with a severity continuum. A considerable proportion of symptoms are related to pathological changes in the vocal system, leading to the assumption that COVID-19 may also affect voice production. For the very first time, the present study aims to investigate voice acoustic correlates of an infection with COVID-19 on the basis of a comprehensive acoustic parameter set. We compare 88 acoustic features extracted from recordings of the vowels /i:/, /e:/, /o:/, /u:/, and /a:/ produced by 11 symptomatic COVID-19 positive and 11 COVID-19 negative German-speaking participants. We employ the Mann-Whitney U test and calculate effect sizes to identify features with the most prominent group differences. The mean voiced segment length and the number of voiced segments per second yield the most important differences across all vowels indicating discontinuities in the pulmonic airstream during phonation in COVID-19 positive participants. Group differences in the front vowels /i:/ and /e:/ are additionally reflected in the variation of the fundamental frequency and the harmonics-to-noise ratio, group differences in back vowels /o:/ and /u:/ in statistics of the Mel-frequency cepstral coefficients and the spectral slope. Findings of this study can be considered an important proof-of-concept contribution for a potential future voice-based identification of individuals infected with COVID-19.
\end{abstract}


\section{Introduction}
\label{sec:introduction}

In December 2019 and early January 2020, a cluster of pneumonia cases with unknown cause emerged in China’s Hubei Province. The pneumonia was found to be caused by a novel coronavirus named \textit{severe acute respiratory syndrome coronavirus 2} (SARS-CoV-2). The disease spread quickly and the first known cases outside of China were identified in mid-January. On 11 February 2020, the World Health Organization (WHO) announced that the disease caused by SARS-CoV-2 would be named COVID-19. A month later, the WHO announced COVID-19 as a pandemic. A year after the emergence of COVID-19, 71\,919\,725 confirmed cases including 1\,623\,064 deaths were reported to the WHO\footnotemark.\footnotetext{\texttt{https://covid19.who.int/}; retrieved 16 December 2020}

The severity of COVID-19 is heterogeneous, ranging from asymptomatic infections or mild flu-like symptoms to severe illness and death. Chest CT \cite{ye2020chest} and post-mortem biopsies \cite{beigmohammadi2020pathological,tian2020pathological} found characteristic pathological changes in patients with COVID-19 and suggest that the lung is the organ that is primarily affected by the disease. Common symptoms of COVID-19 include fever, cough, shortness of breath, weakness, muscle pain, loss of taste and/or smell as well as the ear-nose-throat manifestations sore throat and headache \cite{esakandari2020comprehensive}. Less common ear-nose-throat manifestations of COVID-19 are tonsil enlargement, pharyngeal erythema, nasal congestion, rhinorrhea, and upper respiratory tract infection \cite{el2020ent}. Lechien and colleagues \cite{lechien2020features} reported dysphonia for 26.8\,\% of their investigated patients with mild-to-moderate COVID-19 symptoms. The authors further found a greater severity of COVID-19 symptoms in dysphonic patients compared to non-dysphonic patients. In general, a great proportion of the symptoms associated with COVID-19 affect anatomical correlates of speech production. Important components of the vocal system are the lungs and the lower airway producing the airflow, the vocal folds whose vibrations produce the voice sound, and the vocal and nasal tracts modifying the voice source to produce specific phones, cf.\ \cite{zhang2016mechanics}.

Voice changes have been repeatedly reported for a number of diseases related to pathological changes in components of the vocal system. For example, patients with asthma were found to differ from healthy controls in maximum phonation time (MPT), shimmer, harmonics-to-noise ratio (HNR), jitter, fundamental frequency (F0), first vowel formant (F1), and second vowel formant (F2) \cite{dogan2007subjective,sonu12-DDU}. Singh Walia and Sharma \cite{walia2016level} found the severity of asthma to be related to jitter [\%]. Jitter values derived from recordings of the sustained phonation of the vowel /a:/ was 0.25 for healthy males, 0.41 for males with mild asthma, 0.9 for males with moderate asthma, and 1.83 for males with severe asthma \cite{walia2016level}.
Petrović-Lazić and colleagues \cite{petrovic2011acoustic} reported that jitter, shimmer, F0 variation, voice turbulence index (VTI), pitch perturbation quotient (PPQ), amplitude perturbation quotient (APQ), and HNR values differed between patients with vocal fold polyps and healthy controls. Type and size of vocal fold polyps were found to have effects on jitter and HNR \cite{akbari2018effects}. Male and female patients with unilateral vocal fold paralysis were found to differ from healthy gender-matched controls in jitter, shimmer, HNR, standard-deviation of F0, and standard-deviation of the frequency of F2 \cite{jesus2015acoustic}. In addition, the males with unilateral vocal fold paralysis differed significantly from the male controls in F1 and F2 frequency values as well as in the standard-deviation of the frequency of F1 \cite{jesus2015acoustic}. Segura-Hernández and colleagues \cite{segura2019acoustic} investigated voice characteristics in children with cleft lip and palate before and after speech and language pathology intervention and compared their findings with the voice characteristics of healthy controls. They found that jitter and shimmer were significantly higher in the patients with cleft lip and palate before intervention, whereas the two groups did not differ in these parameters after intervention. In contrast, intervention had no effect on hypernasality, a prominent voice characteristic of patients with cleft lip and palate.

These findings, demonstrating vocal atypicalities in a variety of diseases related to pathological changes in components of the vocal system, lead to the assumption that COVID-19 may be characterised through atypical voice parameters. Characteristic vocal patterns would constitute the starting point for an automatic quick-and-easy-to-apply COVID-19 detection, for example based on smartphone applications. To date, there is hardly any literature on voice parameters of patients with COVID-19. Recently, Asiaee and colleagues \cite{asiaee2020voice} compared voice samples of a sustained vowel /a:/
produced by Persian speakers with and without COVID-19. They extracted the following eight acoustic parameters: F0 and its variations (F0SD), jitter, shimmer, HNR, difference between the first two harmonic amplitudes (H1-H2), MPT, and cepstral peak prominence (CPP). Except F0, all acoustic parameters were significantly different between the patients with COVID-19 and healthy controls. To the best of our knowledge, voice parameters have not yet been analysed for other vowels and there is no study focusing on the voice of German-speaking patients with COVID-19. The present study aims to provide a deeper insight into voice characteristics of patients with COVID-19 by extracting and comparing a comprehensive set of voice parameters from voice samples of the sustained vowels /i:/, /e:/, /o:/, /u:/, and /a:/ produced by German-speaking symptomatic patients with COVID-19 and healthy controls.

\begin{table}[tp]
\centering
\begin{tabular}{r|r|l|r|r}
Vowel & Rank & Feature & \multicolumn{1}{c}{$r$} & \multicolumn{1}{c}{$p$} \\ 
\hline 
\hline
/i:/ & 1 & voiced segments per second	& .46 & .030 \\
 & 2 & mean local shimmer & .43 & .042 \\
 & 3 & mean voiced segment length & .42 & .049 \\
 & 4 & mean rising slope F0 & .41 & .057 \\
 & 5 & rising slope F0 SD & .39 & .066 \\
 & 6 & harmonic difference H1--A3 SD\textsubscript{norm} & .38 & .076 \\
 & 7 & MFCC2 VR SD\textsubscript{norm} & .36 & .088 \\
 & 8 & HNR SD\textsubscript{norm} & .35 & .101 \\
 & 9 & F3 bandwidth SD\textsubscript{norm} & .34 & .115 \\
 & 10 & voiced segment length SD & .33 & .118 \\
 & 11 & F2 bandwidth SD\textsubscript{norm} & .32 & .131 \\
 & 12 & MFCC1 VR SD\textsubscript{norm} & .31 & .149 \\
\hline
/e:/ & 1 & mean voiced segment length & .51 & .017 \\
 & 2 & mean local jitter & .49 & .022 \\
 & 3 & F0 SD\textsubscript{norm} & .48 & .026 \\
 & 4 & voiced segments per second & .48 & .026 \\
 & 5 & HNR SD\textsubscript{norm} & .39 & .066 \\
 & 6 & F3 bandwidth SD\textsubscript{norm} & .38 & .076 \\
 & 7 & MFCC1 VR SD\textsubscript{norm}	& .34 & .115 \\
\hline
/u:/ & 1 & F1 bandwidth SD\textsubscript{norm} & .46 & .030 \\
 & 2 & F3 SD\textsubscript{norm} & .42 & .049 \\
 & 3 & slope\textsubscript{500--1500 Hz} VR SD\textsubscript{norm} & .42 & .049 \\
 & 4 & mean MFCC1 & .39 & .066 \\
 & 5 & MFCC4 SD\textsubscript{norm} & .39 & .066 \\
 & 6 & MFCC1 VR SD\textsubscript{norm} & .39 & .066 \\
 & 7 & mean MFCC1 VR & .38 & .076 \\
 & 8 & loudness pctl\textsubscript{20} & .36 & .088 \\
 & 9 & mean F3 bandwidth & .34 & .115 \\
 & 10 & slope\textsubscript{0--500 Hz} VR SD\textsubscript{norm} & .34	& .115 \\
\hline
/o:/ & 1 & mean F3 bandwidth & .48 & .027 \\
 & 2 & F0 pctlrg\textsubscript{0--2} & .42 & .053 \\
 & 3 & slope\textsubscript{500--1500 Hz} VR SD\textsubscript{norm} & .39 & .073 \\
 & 4 & mean MFCC1 VR & .39 & .073 \\
 & 5 & mean local shimmer & .35 & .113 \\
 & 6 & F3 SD\textsubscript{norm} & .35 & .113 \\
 & 7 & falling slope loudness SD & .33 & .130 \\
 & 8 & mean MFCC1 & .33 & .130 \\
 & 9 & rising slope loudness SD & .32 & .149 \\
 & 10 & mean MFCC2 VR & .32 & .149 \\
\hline
/a:/ & 1 & mean voiced segment length & .39 & .066 \\
 & 2 & mean loudness & .38 & .076 \\
 & 3 & mean MFCC2 VR & .38 & .076 \\
 & 4 & loudness pctl\textsubscript{50} & .35 & .101 \\
 & 5 & spectral flux SD\textsubscript{norm} & .35 & .101 \\
 & 6 & mean MFCC2 & .34 & .115 \\
 & 7 & mean Hammarberg index VR& .32 & .131 \\
 & 8 & loudness pctl\textsubscript{80} & .31 & .149 \\
 & 9 & slope\textsubscript{500--1500 Hz} VR SD\textsubscript{norm} & .31 & .149 \\
 & 10 & loudness peaks per second & .31 & .149 \\
 & 11 & voiced segments per second & .31 & .149 \\
\end{tabular}
\vspace{0.5cm}
\caption{Vowel-wise acoustic features with a differentiation effect $r>.3$ between COVID-19 negative and COVID-19 positive participants, ranked according to the effect size $r$. $r$ is rounded to two decimal places. $p$-values of the underlying Mann-Whitney U tests rounded to three decimal places are given as well. A3 = amplitude of third vowel formant, F0 = fundamental frequency, F1--3 = first to third vowel formant, H1 = relative amplitude of first harmonic, HNR = harmonics-to-noise ratio, MFCC1--4 = first to fourth Mel-frequency cepstral coefficient, pctl = percentile, pctlrg = percentile range, SD\textsubscript{norm} = standard deviation normalised by the arithmetic mean (coefficient of variation), VR = voiced regions}
\label{tab:vowelRanks}
\end{table}

\begin{table}[ht!]
\centering
\begin{tabular}{r|r|l|r|r}
Vowel & Rank & Feature & \multicolumn{1}{c}{$r$} & \multicolumn{1}{c}{$p$} \\ 
\hline 
\hline
/i:/$\cup$/e:/ & 1 & voiced segments per second & .48 & .002 \\
 & 2 & mean voiced segment length & .47 & .002 \\
 & 3 & HNR SD\textsubscript{norm} & .37 & .014 \\
 & 4 & F3 bandwidth SD\textsubscript{norm} & .37 & .014 \\
 & 5 & mean local shimmer & .35 & .022 \\
 & 6 & F0 SD\textsubscript{norm} & .32 & .036 \\
 & 7 & harmonic difference H1--A3 SD\textsubscript{norm} & .31 & .038 \\
 & 8 & MFCC1 VR SD\textsubscript{norm} & .31 & .038 \\
 & 9 & voiced segment length SD & .31 & .039 \\
\hline
/u:/$\cup$/o:/ & 1 & mean F3 bandwidth & .42 & .006 \\
 & 2 & mean MFCC1 VR & .49 & .009 \\
 & 3 & F1 bandwidth SD\textsubscript{norm} & .38 & .012 \\
 & 4 & F3 SD\textsubscript{norm} & .38 & .014 \\
 & 5 & mean MFCC1 & .36 & .018 \\
 & 6 & MFCC1 VR SD\textsubscript{norm} & .34 & .026 \\
 & 7 & mean local shimmer & .33 & .032 \\
 & 8 & F2 bandwidth SD\textsubscript{norm} & .32 & .036 \\
 & 9 & slope\textsubscript{500--1500 Hz} VR SD\textsubscript{norm} & .31 & .040 \\
 & 10 & mean voiced segment length & .31 & .041 \\
 & 11 & mean MFCC2 VR & .30 & .048 \\
\hline
/i:/$\cup$/e:/$\cup$/u:/$\cup$/o:/$\cup$/a:/ & 1 & mean voiced segment length & .39 & $4*10^{-5}$ \\
 & 2 & voiced segments per second & .38 & $8*10^{-5}$ \\
\end{tabular}
\vspace{0.5cm}
\caption{Acoustic features with a differentiation effect $r>.3$ between COVID-19 negative and COVID-19 positive participants, ranked according to the effect size $r$ for the combination of the front vowels /e:/ and /i:/, the back vowels /u:/ and /o:/, and all vowels. $r$ is rounded to two decimal places. $p$-values of the underlying Mann-Whitney U tests rounded to three decimal places are given as well. A3 = amplitude of third vowel formant, F0 = fundamental frequency, F1--3 = first to third vowel formant, H1 = relative amplitude of first harmonic, HNR = harmonics-to-noise ratio, MFCC1--2 = first and second Mel-frequency cepstral coefficient, SD\textsubscript{norm} = standard deviation normalised by the arithmetic mean (coefficient of variation), VR = voiced regions}
\label{tab:vowelRanksComb}
\end{table}

\section{Methods}
\label{sec:methods}
For this study, participants were recruited at the University Medical Center Augsburg (6 patients with COVID-19, 6 healthy controls) and via recruitment flyers from the public (5 patients with COVID-19, 5 healthy controls). Taken together, we include 11 adult participants with COVID-19, referred to as pos.-group, and 11 adult healthy controls, referred to as neg.-group. The participants of the pos.-group and the neg.-group are gender-matched (4 females and 7 males, respectively) and nearly age-matched (pos.-group: mean age = 60 years $\pm$ 20 years standard deviation; age range = 19--79 years; neg.-group: mean age = 55 years $\pm$ 20 years standard deviation; age range = 24--85 years). All participants of the pos.-group were tested positive for COVID-19 within the last 3 days prior to inclusion into the study; all participants of the neg.-group were tested negative for COVID-19 within the last 3 days prior to inclusion into the study. The participants of the neg.-group had neither symptoms of a cold nor chronic pulmonary or voice diseases. The participants of the pos.-group had mild-to-moderate respiratory symptoms of COVID-19. All participants have German as first language and are residents of Germany or Austria. All participants gave their written informed consent for participation in the study. The study procedures are approved by the study commission of the University Medical Center Augsburg, Germany, as well as by the ethics representative of the University of Augsburg, Germany.

The participants were audio-recorded or recorded themselves while producing the sustained vowels /a:/, /e:/, /i:/, /o:/, /u:/ (order like this), representing phonemes of German standard language. They were instructed to make a breathing break after each vowel. Recordings were taken in quiet rooms using a smartphone at a distance of approximately 40 centimeters from the participant's face. In the hospital setting, a Motorola g6 plus smartphone was used. The participants recruited from the public used their own smartphones for the voice recordings.

In a first audio pre-processing step, the recordings are converted into the uniform audio format 16\,kHz/16\,Bit (single channel) PCM by means of FFmpeg\footnotemark.\footnotetext{\texttt{https://ffmpeg.org/}} Then, we use Audacity\footnotemark~to segment the recordings for all single vowels to be exported as separate audio files for the feature extraction step. \footnotetext{\texttt{https://www.audacityteam.org/}}

Acoustic feature extraction is done by means of the open-source toolkit openSMILE\footnotemark~by audEERING\texttrademark~GmbH \cite{Eyben10-OTM,Eyben13-RDI} in its current release 3.0\footnotetext{\texttt{https://github.com/audeering/opensmile}}. 

From each single vowel, we extract the features of the extended Geneva Minimalistic Acoustic Parameter Set (eGeMAPS), representing a compact standard set of 88 acoustic higher-level signal descriptors launched in 2016 by Eyben and colleagues \cite{Eyben16-TGM}. These higher-level descriptors include statistical functionals, such as arithmetic mean, coefficient of variation, percentiles, etc., computed for the trajectories of a range of acoustic time-, energy-, and/or spectral/cepstral-based low-level descriptors, such as F0, Mel-frequency cepstral coefficients (MFCCs), harmonics-to-noise ratio (HNR), jitter, or shimmer. While being a comparably small set among the available openSMILE standard sets, the features of the eGeMAPS were carefully selected by a consortium of engineers, linguists, phoneticians, and clinicians based on their theoretical and practical value for computational voice analysis tasks including clinical applications \cite{Eyben16-TGM}.

We apply the Mann-Whitney U test (group- and vowel-wise feature values are not normally distributed) to analyse the distributions of the extracted acoustic features for differences between the pos.- and the neg.-group. On the one hand, this is done  separately for each vowel. On the other hand, we analyse the combination of the front vowels /i:/ and /e:/, of the back vowels /o:/ and /o:/, as well as of all vowels. To identify the most important acoustic features in either constellation in order to distinguish between individuals from the pos.-group and individuals from the neg.-group, we finally rank the acoustic features according to the effect size $r$ as being the absolute value of Cohen's correlation coefficient \cite{cohen1992statistical}. As null-hypothesis testing with $p$-values as decisive criterion  has been repeatedly criticised (see the statement by the American Statistical Association \cite{Wasserstein16-TAS}, we here report $p$-values as additional descriptive measures and do not employ them for accepting or rejecting a null-hypothesis.

\section{Results}
\label{sec:results}

\begin{figure}[pt]
  \centering
  \includegraphics[width=0.91\textwidth]{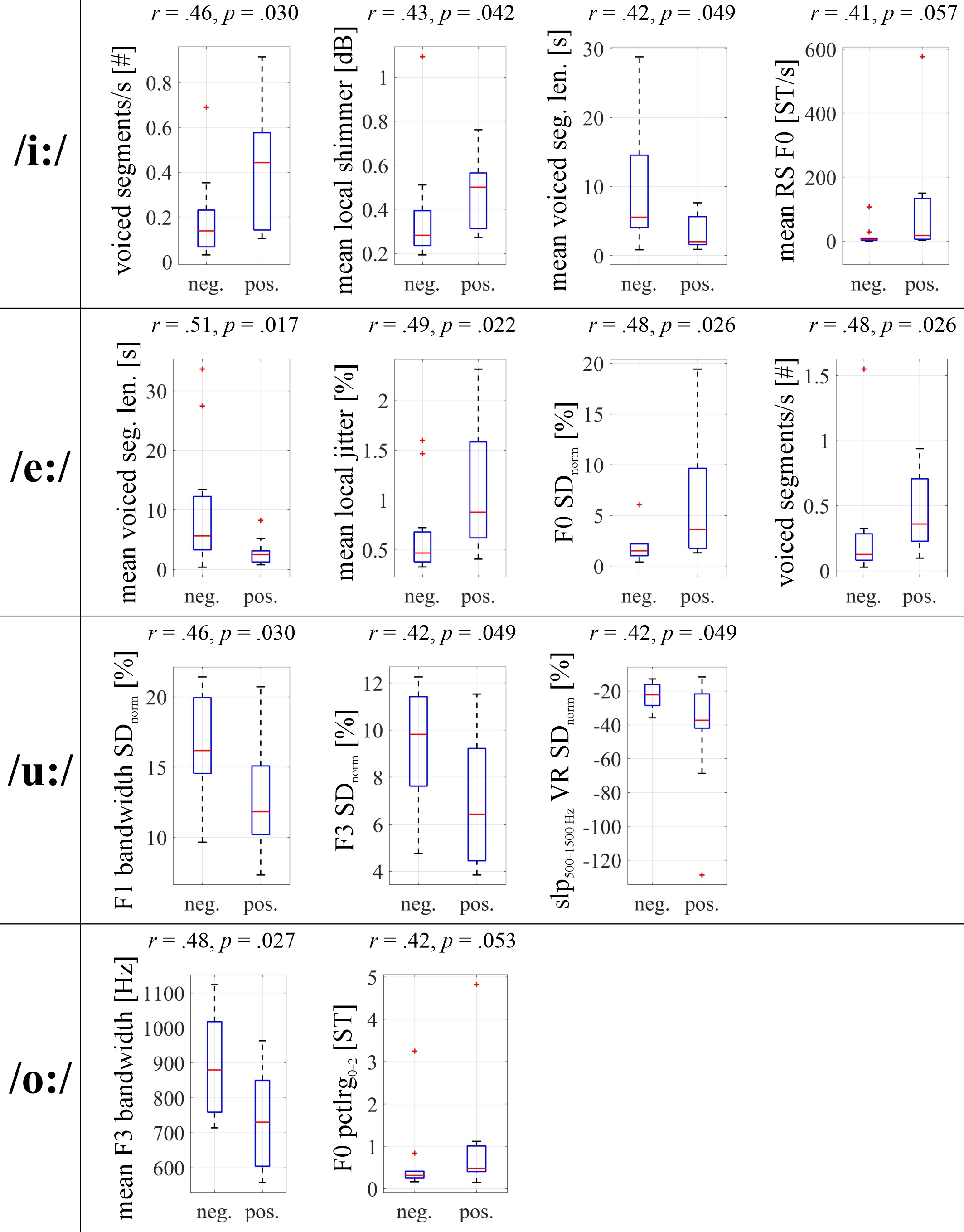}
  \caption{Vowel-wise acoustic feature comparisons between COVID-19 negative (neg.) and COVID-19 positive (pos.) participants in form of boxplots for features with a differentiation effect  $r>.4$ ordered from left to right according to a decreasing $r$, respectively. The effect size $r$ as well as the $p$-value of the Mann-Whitney U difference test are given above each boxplot. $p$ is rounded to three decimal places. $r$ is rounded to two decimal places. Outliers (marked with red plus symbols) are defined as value that are more than 1.5 times the interquartile range away from the bottom or top of the respective box. \# = number of, F0 = fundamental frequency, F1 = first vowel formant, F3 = third vowel formant, len.\ = length, pctlrg = percentile range, RS = rising slope, seg. = segment, ST = semitone from 27.5\,Hz, SD\textsubscript{norm} = standard deviation normalised by the arithmetic mean (coefficient of variation), slp = slope; VR = voiced regions}
  \label{fig:boxplots}
\end{figure}

We display the respective top acoustic features, i.\,e., features differing between the pos.- and the neg.-group with an effect size $r>.3$, for the single-vowel scenarios as well as for the vowel combination scenarios in Tables~\ref{tab:vowelRanks} and \ref{tab:vowelRanksComb}. Additionally, we present boxplots for all features with an effect size $r>.4$ in the single-vowel examinations in Figure~\ref{fig:boxplots}. The mean voiced segment length as well as the number of voiced segments represent the features that differ most between the pos.- and the neg.-group in terms of effect size when combining recordings of all vowels. Further features with prominent group differences across vowels and vowel constellations are bandwidth statistics of the third vowel formant and local shimmer. Additionally, group differences in (a) front vowels are reflected in F0 related statistics and the coefficient of variation of the HNR, and in (b) back vowels in statistics related to the first two MFCCs and in the coefficient of variation of the spectral slope 500--1500\,Hz in voiced regions.

\section{Discussion}
\label{sec:discussion}

In this study, we acoustically analysed sustained vowels produced by participants with a COVID-19 infection and a group of healthy controls. We identified a number of acoustic features to moderately differ between the two groups. We found that the pos.-group produced a higher number of voiced segments per second at a shorter mean voiced segment length as compared to the neg.-group. As participants were instructed to produce sustained vowels, i.\,e., with a continuous phonation over a certain time, this finding may indicate discontinuities in the pulmonic airstream in COVID-19 infected participants leading to sporadic, unintended interruptions of phonation. Voiced segments per second and mean voiced segment length as sort of (overall) duration measures are important in front vowels -- separately and taken together, not amongst the most important in back vowels, and again amongst the important ones in /a:/. They turn out to be most important across all vowels. With a caveat, this might be due to the overall effort that is lower for back vowels whose tongue position is closer to the [\textschwa] (schwa). /a:/ is notoriously prone to `laryngeal irritations'.

Asiaee and colleagues \cite{asiaee2020voice} partly analysed the same acoustic features as used in our study. Among these, the F0 standard deviation, jitter, shimmer, and the HNR were found to be different between COVID-19 positive and COVID-19 negative participants when comparing voice samples of a sustained vowel /a:/. In our study, these features are not among the most important ones to differentiate the groups in the sustained vowel /a:/. However, the normalised F0 standard deviation and local jitter turned out to be relevant for group differentiation in the front vowel /e:/, the normalised HNR standard deviation in the front vowels /i:/ and /e:/, and local shimmer in the vowels /i:/ and /o:/. However, diverging findings between our study and the study by Asiaee and colleagues \cite{asiaee2020voice} may result from the fact that the participants of the latter were Persian speakers, whereas participants in our study are German speakers. All things considered, findings by Asiaee and colleagues \cite{asiaee2020voice} as well as our findings pointing to voice acoustic correlates of a COVID-19 infection across the frequency (e.\,g., F0, formants, jitter), energy (e.\,g., shimmer, HNR), and in our study also spectral/cepstral (e.\,g., MFCCs, slope, harmonic difference) domains suggest that a COVID-19 infection may not be characterisable by a single feature, but by a combination of selected candidate features tied to specific phonation tasks.

Acoustic analysis in this work is based on a compact standard feature set designed for a variety of computational voice analysis tasks also including tasks in clinical context. Starting from the gained knowledge about each feature's relevance for reflecting vocal differences between COVID-19 positive and COVID-19 negative speakers, future work should additionally focus on specific clinical speech parameters that allow for interpretations from a voice-physiological point of view, such as the glottal-to-noise excitation (GNE) ratio \cite{michaelis97-GNE}.

A limitation of our study is the relatively small sample size. Furthermore, the neg.-group consists of healthy speakers only, i.\,e., speakers without any symptoms of a cold, whereas the pos.-group only includes patients with mild-to-moderate flu-like symptoms. To evaluate whether there are voice parameters specific for a COVID-19 infection, future studies need to include a considerable amount of patients with COVID-19 who do not show respiratory or ear-nose-throat symptoms and COVID-19 negative participants with cold-like symptoms. Moreover, as some of the acoustic features we show to be important for a differentiation between COVID-19 positive and COVID-19 negative participants were also reported to be relevant for differentiating between patients with asthma and healthy controls \cite{dogan2007subjective,sonu12-DDU,walia2016level}, it is highly important for future studies to include patients with asthma and other chronically ill patients in a control group.

Despite the limitations, our study can be regarded as a first step towards unravelling the complex acoustic fingerprint of COVID-19 and as an important proof-of-concept achievement for future voice-based viral infection identification applications. A re-validation of our findings based on a much larger and more heterogeneous sample is warranted.

\section*{Acknowledgement}

The authors express their gratitude to all participants for `donating' their voices though facing challenging times by being directly or in any case indirectly affected by the COVID-19 pandemic. The authors want to thank them for being open-minded for innovative research approaches promoting progress in science and global healthcare.    

\bibliographystyle{unsrt}  
\bibliography{references}

\begin{thebibliography}{10}

\bibitem{ye2020chest}
Zheng Ye, Yun Zhang, Yi~Wang, Zixiang Huang, and Bin Song.
\newblock Chest {CT} manifestations of new coronavirus disease 2019
  {(COVID-19)}: a pictorial review.
\newblock {\em European Radiology}, 30:4381–4389, 2020.

\bibitem{beigmohammadi2020pathological}
Mohammad~Taghi Beigmohammadi, Behnaz Jahanbin, Masoomeh Safaei, Laya Amoozadeh,
  Meysam Khoshavi, Vahid Mehrtash, Bita Jafarzadeh, and Alireza Abdollahi.
\newblock Pathological findings of postmortem biopsies from lung, heart, and
  liver of 7 deceased {COVID-19} patients.
\newblock {\em International Journal of Surgical Pathology}, 2020.

\bibitem{tian2020pathological}
Sufang Tian, Yong Xiong, Huan Liu, Li~Niu, Jianchun Guo, Meiyan Liao, and
  Shu-Yuan Xiao.
\newblock Pathological study of the 2019 novel coronavirus disease {(COVID-19)}
  through postmortem core biopsies.
\newblock {\em Modern Pathology}, 33:1007–1014, 2020.

\bibitem{esakandari2020comprehensive}
Hanie Esakandari, Mohsen Nabi-Afjadi, Javad Fakkari-Afjadi, Navid Farahmandian,
  Seyed-Mohsen Miresmaeili, and Elham Bahreini.
\newblock A comprehensive review of {COVID-19} characteristics.
\newblock {\em Biological Procedures Online}, 22:1--10, 2020.

\bibitem{el2020ent}
Mohammad~Waheed El-Anwar, Saad Elzayat, and Yasser~Ahmed Fouad.
\newblock {ENT} manifestation in {COVID-19} patients.
\newblock {\em Auris Nasus Larynx}, 47(4):559--564, 2020.

\bibitem{lechien2020features}
Jerome~R Lechien, Carlos~M Chiesa-Estomba, Pierre Cabaraux, Quentin Mat, Kathy
  Huet, Bernard Harmegnies, Mihaela Horoi, Serge~D Le~Bon, Alexandra Rodriguez,
  Didier Dequanter, et~al.
\newblock Features of mild-to-moderate {COVID-19} patients with dysphonia.
\newblock {\em Journal of Voice}, 2020.

\bibitem{zhang2016mechanics}
Zhaoyan Zhang.
\newblock Mechanics of human voice production and control.
\newblock {\em Journal of the Acoustical Society of America},
  140(4):2614--2635, 2016.

\bibitem{dogan2007subjective}
Muzeyyen Dogan, Emel Eryuksel, Ismail Kocak, Turgay Celikel, and Mehmet~Ali
  Sehitoglu.
\newblock Subjective and objective evaluation of voice quality in patients with
  asthma.
\newblock {\em Journal of Voice}, 21(2):224--230, 2007.

\bibitem{sonu12-DDU}
Sonu and R.~K Sharma.
\newblock Disease detection using analysis of voice parameters.
\newblock {\em International Journal of Computer Science and Communication
  Technology}, 4(2):4--6, 2012.

\bibitem{walia2016level}
Gursimarjot~Singh Walia and RK~Sharma.
\newblock Level of asthma: mathematical formulation based on acoustic
  parameters.
\newblock In {\em 2016 Conference on Advances in Signal Processing (CASP)},
  pages 24--27. IEEE, 2016.

\bibitem{petrovic2011acoustic}
Mirjana Petrovi{\'c}-Lazi{\'c}, Sne{\v{z}}ana Babac, Mile Vukovi{\'c}, Rade
  Kosanovi{\'c}, and Zoran Ivankovi{\'c}.
\newblock Acoustic voice analysis of patients with vocal fold polyp.
\newblock {\em Journal of Voice}, 25(1):94--97, 2011.

\bibitem{akbari2018effects}
Elaheh Akbari, Sadegh Seifpanahi, Ali Ghorbani, Farzad Izadi, and Farhad
  Torabinezhad.
\newblock The effects of size and type of vocal fold polyp on some acoustic
  voice parameters.
\newblock {\em Iranian Journal of Medical Sciences}, 43(2):158, 2018.

\bibitem{jesus2015acoustic}
Luis~MT Jesus, Joana Martinez, Andreia Hall, and An{\'\i}bal Ferreira.
\newblock Acoustic correlates of compensatory adjustments to the glottic and
  supraglottic structures in patients with unilateral vocal fold paralysis.
\newblock {\em BioMed Research International}, 2015:704121, 2015.

\bibitem{segura2019acoustic}
Monica Segura-Hern{\'a}ndez, V{\'\i}ctor~Manuel Valadez-Jim{\'e}nez,
  Pablo~Antonio Ysunza, Araceli~Patricia S{\'a}nchez-Valerio, Emilio
  Arch-Tirado, Ana~Luisa Lino-Gonz{\'a}lez, and Xochiquetzal
  Hern{\'a}ndez-L{\'o}pez.
\newblock Acoustic analysis of voice in children with cleft lip and palate
  following vocal rehabilitation. preliminary report.
\newblock {\em International Journal of Pediatric Otorhinolaryngology},
  126:109618, 2019.

\bibitem{asiaee2020voice}
Maral Asiaee, Amir Vahedian-Azimi, Seyed~Shahab Atashi, Abdalsamad Keramatfar,
  and Mandana Nourbakhsh.
\newblock Voice quality evaluation in patients with {COVID-19}: An acoustic
  analysis.
\newblock {\em Journal of Voice}, 2020.

\bibitem{Eyben10-OTM}
Florian Eyben, Martin W\"ollmer, and Bj\"orn Schuller.
\newblock {openSMILE: The Munich versatile and fast open-source audio feature
  extractor}.
\newblock In {\em Proceedings of the 18\textsuperscript{th} ACM International
  Conference on Multimedia, MM 2010}, pages 1459--1462, Florence, Italy,
  October 2010. ACM.

\bibitem{Eyben13-RDI}
Florian Eyben, Felix Weninger, Florian Gro{\ss}, and Bj\"orn Schuller.
\newblock {Recent developments in openSMILE, the Munich open-source multimedia
  feature extractor}.
\newblock In {\em Proceedings of the 21\textsuperscript{st} ACM International
  Conference on Multimedia, MM 2013}, pages 835--838, Barcelona, Spain, October
  2013. ACM.

\bibitem{Eyben16-TGM}
Florian Eyben, Klaus R.Scherer, Bj\"orn~W. Schuller, Johan Sundberg, Elisabeth
  Andr{\'e}, Carlos Busso, Laurence~Y. Devillers, Julien Epps, Petri Laukka,
  Shrikanth Narayanan, and Khiet~P. Truong.
\newblock {The Geneva Minimalistic Acoustic Parameter Set (GeMAPS) for voice
  research and affective computing}.
\newblock {\em IEEE Transactions on Affective Computing}, 7(2):190--202, 2016.

\bibitem{cohen1992statistical}
Jacob Cohen.
\newblock Statistical power analysis.
\newblock {\em Current Directions in Psychological Science}, 1(3):98--101,
  1992.

\bibitem{Wasserstein16-TAS}
Ronald~L. Wasserstein and Nicole~A. Lazar.
\newblock {The ASA's statement on p-values: context, process, and purpose}.
\newblock {\em The American Statistician}, 70(2):129--133, 2016.

\bibitem{michaelis97-GNE}
Dirk Michaelis, Tino Gramss, and Hans~Werner Strube.
\newblock Glottal-to-noise excitation ratio--a new measure for describing
  pathological voices.
\newblock {\em Acta Acustica united with Acustica}, 83(4):700--706, 1997.

\end{thebibliography}

\end{document}